\documentclass{IETpaper}
\usepackage{graphicx}
\usepackage{balance}

\begin{document}


\title{If you can't understand it, you can't properly assess it! \\The reality of assessing security risks in \\Internet of Things systems}

\author{Jason R C Nurse$^{\ast}$, Petar Radanliev$^{\dag}$, Sadie Creese$^{\ast}$, David De Roure$^{\dag}$}

{\centering
\institution{$^{*}$Department of Computer Science, University of Oxford, UK, 
$^{\dag}$Oxford e-Research Centre, University of Oxford, UK\\
jason.nurse@cs.ox.ac.uk, petar.radanliev@oerc.ox.ac.uk, sadie.creese@cs.ox.ac.uk, 
david.deroure@oerc.ox.ac.uk
}
}

\maketitle


\begin{abstract}
Security risk assessment methods have served us well over the last two decades. As the complexity, pervasiveness and automation of technology systems increases, particularly with the Internet of Things (IoT), there is a convincing argument that we will need new approaches to assess risk and build system trust. In this article, we report on a series of scoping workshops and interviews with industry professionals (experts in enterprise systems, IoT and risk) conducted to investigate the validity of this argument. Additionally, our research aims to consult with these professionals to understand two crucial aspects. Firstly, we seek to identify the wider concerns in adopting IoT systems into a corporate environment, be it a smart manufacturing shop floor or a smart office. Secondly, we investigate the key challenges for approaches in industry that attempt to effectively and efficiently assess cyber-risk in the IoT.
\end{abstract}

\section{Introduction}
\label{Introduction}
The Internet of Things (IoT) is set to change our society in ways potentially as significant as the internet itself. Beyond the buzzword and supposition of all devices being connected, the IoT is actually a complex technological paradigm. This is, in part, due to the reality that it represents the convergence of varying visions for the future of technology~\cite{atzori2010internet}. The disruptive nature of the IoT presents society with a range of advantages but also several noteworthy challenges with its widespread use. 

On the one hand, the IoT stands to significantly increase productivity and efficiency in domains such as manufacturing and agriculture. Some studies suggest that the expected economic impact is likely to reach at least \$4 trillion per year by 2025~\cite{technative2017}. Apart from economics, there is the tangible likelihood of this new paradigm to save lives when we reflect on the domain of smart health~\cite{miorandi2012internet}, for example. On the other hand however, security in such a disruptive and complex paradigm can prove extremely challenging. Threats may originate from physical or cyber-attacks and may target any of the central layers of an IoT system; many of which are known to have shortcomings in terms of security and privacy~\cite{kumar2016security}.

The way that organisations have been conditioned to respond to threats and vulnerabilities in systems --- be they IoT-oriented or otherwise --- is guided by the process of risk assessment. Such assessments often occur through the use of well-regarded methodologies such as the National Institute of Standards and Technology's (NIST) Special Publications (e.g., SP800-30, SP800-37), industry-developed standards including the ISO 27000 series, and others. These assessments have the goal of identifying relevant assets, vulnerabilities and threats, along with associated likelihoods and impacts; when appropriately combined, these then lead to the definition of risks facing a system. These risk assessment approaches have served us well over the past decade, and have provided a platform through which organisations and governments could better protect themselves against pertinent risks. 

There are, however, several issues which we believe will impact the application of existing security assessment methodologies to IoT systems. These include the inadequate nature of current periodic (e.g., quarterly or annually) assessments, unknown system boundaries at the time of assessment given dynamic IoT systems, and failure to consider assets as avenues of attack instead of only as items of value~\cite{nurse2017security}. These are all key issues which we posit raise substantial challenges for organisations as they aim to assess risk in IoT or connected systems, before then deciding on appropriate risk treatments.

In this paper we build on our earlier research~\cite{nurse2017security} by reporting on a series of scoping workshops and interviews with professionals from industry and business sectors. The aim of this engagement has been to validate the key issues identified above and their significance as it pertains to identifying and assessing cyber-risk. We view this as a novel and essential contribution because it ensures that our research is well-informed by current practice, and real business scenarios and context. Furthermore, we believe that the IoT security research field in general stands to benefit from this stakeholder engagement. The reason for this is because it aids in elucidating the real concerns held by industry, and informing the direction of future research. 

Our article is structured as follows. Section~\ref{Related work} briefly reflects on related research in the domain of cyber-risks, risk assessment for systems and connected IoT systems. Next we present an overview of the core issues that are hypothesised to complicate IoT risk assessment in Section~\ref{Assessing security risks in IoT systems}.  Section~\ref{Consulting experts from industry} then details the research approach adopted to examine the aforementioned issues and related cyber security challenges with industry professionals. This is followed by Section~\ref{Findings and discussion} which presents and discusses the core research findings from the study. Here we also highlight the implications for IoT risk assessment research going forward. The article then concludes in Section~\ref{Conclusion and future work} and defines avenues for future work.

\section{Background and related work}
\label{Related work}

\subsection{IoT systems and security concerns}

The Internet of Things is the intersection of three visions for the future of technology~\cite{atzori2010internet}. These are the Things-oriented vision (essentially, the use and presence of various electronically tagged things), the Semantic-oriented vision (meant to address issues of how to represent, connect and store items) and the internet-oriented vision (which encourages the use of web standards to interconnect items). Hints of each of these visions can be found in commonly used definitions for IoT, i.e., interconnected networks of digitised physical devices which interact to achieve some purpose. 

IoT systems can generally be divided into the following environments: applications, cloud services and things (physical or digital)~\cite{nurse2017security}. Applications chart the objective of the IoT system, and cover domains such as smart health, smart factories and building automation. The purpose of the cloud environment (cloud computing or services) is to compose and enact a series of dynamic services (typically software components) to realise the application. Things are used by services to interact with the real world, and include devices, sensors and actuators. These can be called upon, changed or added to as needed. 

Since its inception, security, privacy and trust have been key concerns in IoT systems. Security, in particular, has drawn significant attention given its use as an enabler for achieving privacy and system trust. There are many challenges to attaining security in the IoT, but some of the most central in the literature pertain to identity and authentication, access control, protocol and network security, fault tolerance and governance~\cite{kumar2016security,roman2013features}. For any implementation of IoT to be successful, these need to be adequately researched and addressed. 

At the more granular level, another area covered by related works is the definition of risks and threats from the integration of IoT into systems. In Babar \textit{et al.}~\cite{babar2010proposed} for instance, an IoT threat taxonomy has been proposed which defines threats to storage management (e.g., key management confidentiality), communication (e.g., denial of service on IoT devices), dynamic binding (issues regarding naming and addressing of connected things) and embedded systems (e.g., side-channel and tampering attacks). Other work outlines a layered approach to understanding IoT security issues, with threats and solutions defined in terms of the application, transport and perception layers~\cite{jing2014security}. A notable observation from this reflection is that many existing security issues are exacerbated by the IoT context---these may be general problems such as Distributed Denial-of-Service (DDoS) attacks or specific issues including insider threat~\cite{nurse2015smart,krebsdyn2016}. A key factor in these cases is the low-resource nature of IoT devices, their pervasiveness, and their open accessibility over the internet. 

\subsection{The risk assessment context}

To manage security concerns, organisations typically rely on some form of risk management process. Within risk management, risk assessment allows for the identification and prioritisation of risks (and inclusive factors such as threats, vulnerabilities, impacts, etc.), while risk treatment considers the enterprise's security posture before determining how to treat each risk defined. A few common examples of approaches that are used and promoted to assess risks are NIST SP800-30~\cite{nistsp800}, OCTAVE~\cite{cmuoctave2017}, IRAM2~\cite{isf2017} and ISRAM~\cite{karabacak2005isram}. Overall, these follow similar underlying methods and only differ in how they orient themselves, for instance, around assets/threats or using qualitative or quantitative assessment ratings.

For the IoT context however, there are few methods proposed in the literature to assess IoT system risks. In many cases, traditional methods (such as those above) are applied to IoT scenarios or the general guidance put forward is not tailored to IoT systems or their dynamics; see~\cite{enisa2016}. This is quite concerning as we will discuss in Section~\ref{Assessing security risks in IoT systems} when we reflect on the shortcomings of such approaches. Of the methods specifically created for the IoT, we noticed that these tended to focus more on automated, mathematical approaches. This was as opposed to process-driven techniques similar to NIST, OCTAVE, and the others mentioned above. 

The most noteworthy of the IoT approaches include: the IoTRiskAnalyzer framework which formally and quantitatively analyses IoT risks using probabilistic model checking~\cite{mohsin2017iotriskanalyzer}; the framework proposed by Ge \textit{et al.} for graphically modelling and assessing security for the IoT through formal system definitions~\cite{ge2017framework}; methods which adopt Bayesian techniques to assessment including attack graphs and inference networks~\cite{wu2014novel,munoz2016bayesian}; and SecKit, a model-based security toolkit for identifying and addressing IoT risks~\cite{neisse2015seckit}.  These methods generally seek to provide an automated way to conduct risk assessment and thereby increase efficiency while removing some subjectivity from the traditional manual process. 

\section{Assessing security risks in IoT systems}
\label{Assessing security risks in IoT systems}

Risk assessment is a difficult process in general~\cite{engel2017risk,taylor2015potential}, and one which we believe is even more challenging when assessing risks in IoT systems. This is particularly the case when traditional approaches to assess risk are applied (as is largely the case today), because they fail to cater for the nuances of the IoT. In our prior research~\cite{nurse2017security}, we have outlined four key reasons why security risks assessments as currently designed are lacking for the IoT context. We briefly recap these below.

The first concern is that current risk assessment approaches are based on periodic assessment and assume that systems will not significantly change in a short period of time. These assumptions do not hold for the IoT, where there is vast variability in scale of systems, dynamism and system coupling. 

\begin{figure}[ht]
	\centering
	\includegraphics[scale=0.55]{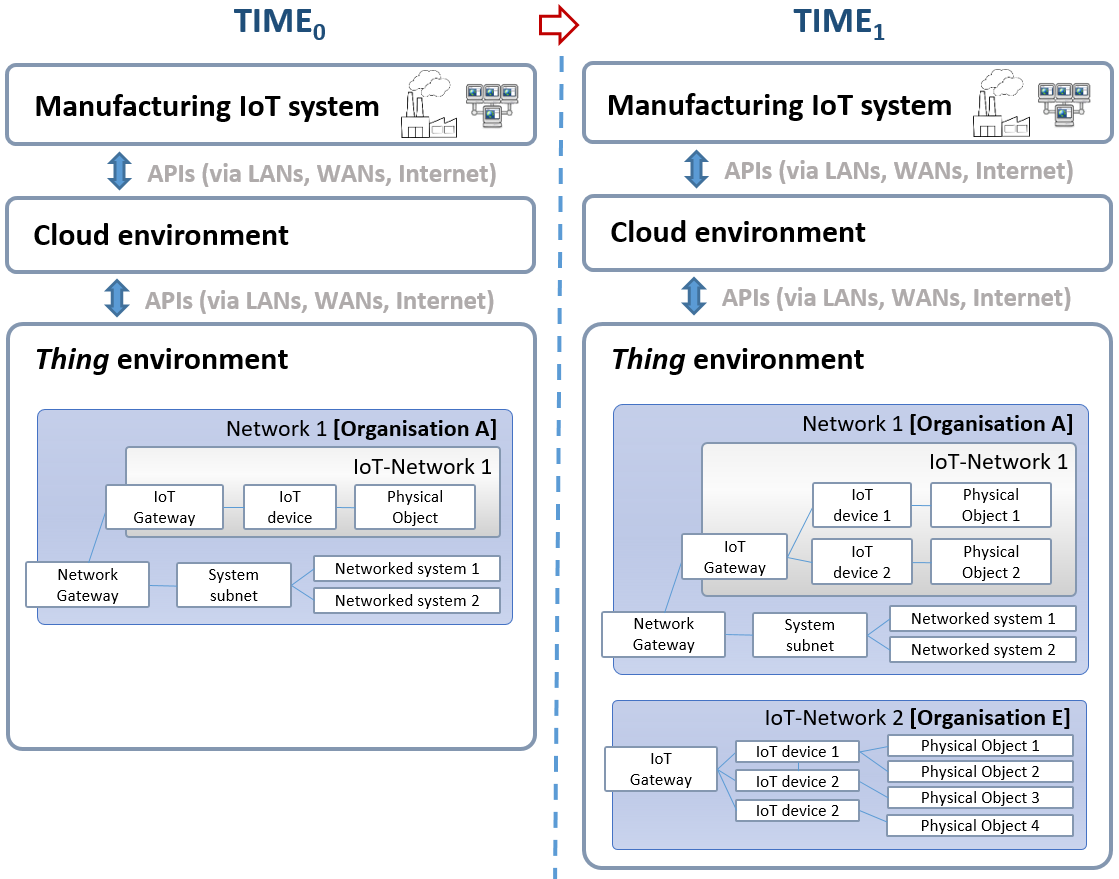}
	\caption{IoT systems dynamically changing over time}
	\label{figure:timechange}
\end{figure}

If we take Figure~\ref{figure:timechange} as an example, at $Time\textsubscript{0}$ the manufacturing IoT system may be composed of a specific set of services and things. At $Time\textsubscript{1}$ (which may be hours or days later) however, we can see that this set has increased based on the items in the Things environment. This may be due to changing needs of the system, adaptations to increase efficiency or newly interconnected services---a new third party (Organisation E) may needed to provide machine parts or support in-situ systems for instance. The difficulty faced is that a risk assessment may be conducted at $Time\textsubscript{0}$ but the adaptation at $Time\textsubscript{1}$ may arise long before the system is due for reassessment (which is typically a quarterly or bi-annual activity in organisations). The traditional processes, therefore, can lead to drastically outdated assessments.

A second factor to consider is that the detailed knowledge typically required to conduct traditional risk assessments (on assets, threats, vulnerabilities, and so on) is extremely challenging to attain considering the highly dynamic nature of IoT systems. Systems are difficult to keep track of and often have unknown components a priori~\cite{atzori2010internet}. In our Figure~\ref{figure:timechange} example, at $Time\textsubscript{0}$ risk assessors may have little knowledge of the full complement of systems that may be integrated at runtime (i.e., the extent of $Time\textsubscript{1}$). Without a good understanding of the assets constituting the system, assessors cannot use existing methods to reason about potential threats, vulnerabilities, attack likelihoods or impacts. Even in cases where this data exists, it may be internal to a partnering company (e.g., Organisation E) not willing to share it and risk exposing themselves.

Another concern is that traditional assessments focus on tangible assets often at the expense of intangibles. This problem has been hinted at in existing research where tacit knowledge---which in some cases is more important than physical assets---was found to be overlooked in existing security risk assessments~\cite{shedden2011incorporating}. In the IoT, such oversights are concerning because of the presence of key intangibles including the processes through which devices are bound, the connections that allow them to couple and operate, and the inner workings of the system actors themselves. Each of these is a potential area of new risk which needs to be adequately risk assessed. As an example, consider the case where attackers are able to gain insight into the logical setup of cloud services or how IoT assets were dynamically referenced and bound. This could be used to determine where best to target to disrupt the overall application. This however, is not a focus of standard assessment techniques.

Lastly, as was witnessed in the 2016 Dyn cyber-attack~\cite{krebsdyn2016} which leveraged compromised IoT devices (much like those in the Things environment in Figure~\ref{figure:timechange}), there is a strong argument for viewing organisational assets not only as items of value but as attack platforms. This view is supported by earlier works such as the intrusion kill chain, where compromised assets can be used to further attacks~\cite{hutchins2011intelligence}. It would be prudent, therefore, for organisations to accommodate for these new types of risks into their risk assessment processes. This currently is not done, and it raises numerous new questions pertaining to where the boundaries of risk assessment in the IoT should lie. 

\section{Our approach to consulting industry experts}
\label{Consulting experts from industry}


While academic literature and industry reports can provide useful insight into the challenges encountered when organisations try to assess risks in connected and IoT infrastructures, there is also great value in consulting industry experts directly. Such individuals may be used to validate (or invalidate) the issues identified, and further elucidate the peculiarities of risk assessment in connected systems.
To achieve this, we reached out to several industry-based professionals with experience, knowledge and expertise in the areas of enterprise systems, IoT, engineering, risks assessment, and organisational security. This followed a snowball sampling approach where we polled our direct contacts and sought to recruit others via those contacts. We then filtered the respondents based on profession and experience to ensure that they had the necessary expertise. 

Once the set of professionals were recruited, we then conducted a series of small workshops and interviews. Workshops were operated similar to focus groups~\cite{berg2004qualitative}, and therefore allowed us to pose discussion points (guided questions) and then step back whilst professionals deliberated the topic. We used semi-structured interviews for participants unable to attend workshops. The semi-structured format was preferred to facilitate guided questioning (identical to the workshops) but also to allow for unplanned avenues of inquiry to be pursued. Subsequently, all data gathered was transcribed and we applied the content analysis technique~\cite{berg2004qualitative} to assess and draw insights from our findings. 

Considering that a key aim of the study was to validate or refute the issues identified in our earlier work, we decided to prepare a briefing document for circulation before the study. This brief contained an expanded version of the four core issues put forward in Section~\ref{Assessing security risks in IoT systems}. We asked participants to critically reflect on the document before the study, and for all cases, it was sent two weeks prior to the session. 

The flow of the sessions involved first gathering professionals' general feedback (agreement or disagreement) on the points raised in the brief and then posing a series of questions specifically pertaining to cyber-risk in IoT systems. These aimed to elicit professional views regarding IoT embedded into corporate environments, perspectives on risk assessment generally, and any pressing concerns they had regarding risk assessment given the dynamics of the IoT. To summarise a number of the key points, we also presented a set of hypotheses for participant feedback. These included: 

\vspace{-1em}
\begin{itemize}
	\itemsep0em
	\item Risk assessment approaches for the IoT that adopt a periodic assessment instead of incorporating the potential for changes in the IoT system (e.g., shifting boundaries) will miss significant risks.
	\item Limited system knowledge in IoT and coupled systems is a key issue impacting security risk assessment.
	\item The logical glue through which IoT systems are bound is an area not adequately covered in current risk assessment, but if it is exploited, it could have a significant impact on IoT systems.
	\item Risk assessments that only consider assets as items of value as opposed to items that may be used as an attack platform, will miss risks in the IoT environment.  
\end{itemize}

Having introduced the study approach, next we present and discuss the findings.

\section{Results and discussion}
\label{Findings and discussion}

The aim of our study was to gain an in depth understanding of the challenges that professionals face as they assess risks in environments containing IoT systems. Following our recruitment and study structure, we conducted two workshops and one interview, most lasting one hour. All participants were knowledgeable in enterprise systems, risk assessment and IoT systems, and were employed by medium-to-large technology enterprises. In the sections that follows, we identify and reflect on the most significant points emerging from the study. 

\subsection{Periodic assessment versus an evolving IoT system}

In our discussions with industry professionals, one of the most noteworthy points to emerge was that IoT systems, by their nature, are capable of continuously evolving during operations. This evolution encompasses the addition of new devices and services (of varying levels of intelligence and processing power) to the system, and the automated adaptation of the system to cater to the application scenario. Such new devices may include sensors intended to take measurements of some physical object or tertiary cloud services instantiated to support reasoning and prediction of actions based on those new measurements. As a result of this constant evolution, the study participants expressed that risk assessment is substantially more challenging and that periodic assessment would not be sufficient.

One practitioner aptly summarised the others, while also making specific reference to the scale of devices and resulting difficult in attempting to assess risks. During his workshop, he stated: 

\begin{quote}
\textit{..because of the explosion in devices, it is difficult to keep track of these devices and their capabilities ... Knowing which devices are potentially compromisable and which are not, is an ongoing challenge, and one that changes on a day-by-day basis.}
\end{quote}

The central point emerging here was that because IoT systems can change so quickly, periodic assessments stand to overlook the increasing variety of risks that accompany these newly added devices, services and technologies. This speaks directly to our first hypothesis and supports its validity. The issue of `keeping up with risks' (due to a difficulty in `keeping up with the system assets') was seen as one of the primary predicaments and a core shortcoming of current risk assessment systems when applied to any dynamic system context.

In addition to the challenges above, participants identified two other issues with periodic assessment which are likely to be exacerbated when applied to IoT systems; both of these relate to the security culture of organisations. The first issue pertained to the fact that in many organisations today, particularly  Small-to-Medium-sized Enterprises (SMEs), security is viewed as a single point to be reached as opposed to continuous set of actions and behaviours. As commented by one professional: 

\begin{quote}
	\textit{..once they have reached a level of security, they are often tempted to think that the task is complete. The fact that security is dynamic and is changing all the time is something these companies will need to get accustomed to as this is not their expectation.}
\end{quote}

This is an important observation because single levels of security are often supported by, and suited to, periodic risk assessments. That is, once an assessment is complete and security measures have been implemented, the security problem is regarded as `solved' for another year. These perceptions view systems and security as static items, while the reality is that maintaining IoT systems and bolstering their security is now a more dynamic task than before. With regards to the emphasis on SMEs above, this might be expected given their limited budgets and confirmed reports into underinvestment in cybersecurity~\cite{scmag2017smes}. The reality is, however, that SMEs often work with larger corporations and internal risks can easily be passed from one to the other. 

The second issue supports the first point and highlights the belief of a few participants that some companies still view risk assessment as a checkbox exercise. In those cases, risk assessment is not motivated by a real desire to identify, prioritise and address risk and ensure organisations are adequately protected. Instead, it is driven by regulations and process. Although the link between security and regulations is well-known~\cite{dcmscsr2016}, regulations may be years behind systems as progressive as the IoT. A good security culture which appreciates  risk is paramount for systems today, and especially for the connectivity facilitated in coupled systems.

\subsection{Risk assessing a complex, autonomous black box}

Another area of focus within our sessions with industry professionals was the level of knowledge and insight typically maintained by users and security teams into IoT systems. We were interested in understanding this generally, but especially in determining how knowledge (or lack thereof) has an impact on how risk are assessed in organisations. The first salient point made by the group was that IoT systems are becoming extremely complex and have constantly shifting system boundaries---these factors, according to the participants, make it increasingly difficult to do good risk assessment. Speaking on this situation, a participant stated:

\begin{quote}
	\textit{If you can't understand it [the system], you can't properly assess it! Also, failure to understand a system means that if people make changes to it, it is not possible to appreciate the implications. This is a common problem in software engineering when trying to make changes in complex software systems.}
\end{quote}

This perspective identifies the difficulty in risk assessment due to system complexity, but also highlights the real challenge of tracking changes and updates to the system and their implications. This is a salient observation because it suggests that knowing the devices and services present within an IoT system is not sufficient. We also need to understand and track how changes are made to the underlying systems, their impact on the wider system, and the subsequent repercussions on risk. 

While intelligent, highly dynamic, inter-organisational systems are a fundamental part of the IoT, participants also felt that these features introduced numerous issues in defining and tracking risks. In some highly automated and intelligent IoT systems for instance, machines reason and communicate without the involvement of individuals, which means that in many cases workers may not fully understand the spectrum of risks in such systems. This is particularly problematic from a legal perspective if issues arise and there is a requirement to determine who is responsible. There is also the real situation that developers of original IoT systems may be employees who have since left the organisation or external firms who offer limited support. Both of these cases can result in a severely limited understanding of enterprise IoT risk and where that risk resides.

A related point worthy of mention is that according to some professionals, IoT systems are black boxes. This perception is caused by the difficulty in knowing exactly what is happening on the system or its various connected components. The limited knowledge of these new systems is especially concerning because, as participants mentioned, they are connected to the Internet and thereby open a new platform of attack (be it infiltration or exfiltration). Additionally, professionals expressed concern about individual smart devices being integrated into corporate networks---these can span from sensors on manufacturing floors to office smart displays or Amazon Alexa for work. As one professional summarised:

\begin{quote}
	\textit{Traditionally, companies would have full control over systems on their networks and be able to secure them and interact as desired. ... With IoT, devices are now placed on organisational networks but they are complete `black boxes'. You have no control over the software or what the device is doing, other than what is disseminated by the manufacturer/provider. }
\end{quote}

This opinion touches on the issues of control and trust, which have become core to the success of connected systems. In the IoT, organisations are often required to relinquish control and trust other parties. This is contrary to how risk assessment approaches today function; these typically require a full complement of data on systems and assets, related vulnerabilities and threats. These complex, autonomous black boxes, as described by participants, therefore pose a fundamental challenge to how we think about risk assessment for IoT systems.

\subsection{How to address new elements of IoT risk?}

The topic of new elements of risk in IoT systems was also posed to participants. Specifically, we wanted to understand two aspects related to our earlier hypotheses. Firstly, we aimed to determine the extent that existing risk assessments considered the risk related to the logical glue (or innate knowledge) that binds systems. Secondly, our goal was to gather professionals' views on whether assessments may be lacking by only regarding assets as items of value instead of also as platforms of attack.

Participants views on these points were less detailed than other issues, nonetheless, most individuals felt that both were valid concerns worthy of some consideration in IoT risk assessment. With regards to the binding of systems for instance, professionals could see how an intimate knowledge of how a system works, even if only at the process-level, could lead to the introduction of new risk if misused. What was not clearly understood by interviewees however, was how such risks would be addressed in current assessment techniques. These techniques appeared to be suited towards risks which focused on tangible enterprise assets or threats. A similar situation emerged when exploring the second point about viewing assets as platforms of attack. To quote a participant:

\begin{quote}
	\textit{The risk assessment has traditionally been the first line of defence. It is, how to stop being compromised, not, if you are compromised, how does that then propagate i.e., what's the next stage of the attack.}
\end{quote}

This provides valuable insight into how some professionals view the scope of risk assessment, i.e., as only the initial step. When we explored this point further to determine what corporate security measures may cater for assets as attack platforms, attack and impact pathways were mentioned by an interviewee. These were regarded as more `live' than risk assessments and would model how the compromise of one system may lead to other systems being exposed. Even in this case however, the practitioner noted that it quickly becomes difficult to assess all the possible impacts of something occurring. This difficulty in traditional systems will only be exacerbated in the IoT, regardless of if these pathways are explored within or outside of the risk assessment process.

\subsection{Automated risk assessment in the IoT: Yea or Nay}

The notion of automated risk assessment also led to significant discussion amongst practitioners. This pertained to the question of whether automated risk assessment was a feasible solution given how dynamic the IoT can become. Overall, participants expressed that a fully automated risk assessment process for the IoT was not feasible because of the presence of social and human aspects in systems. This refers to the difficulty of modelling and unpredictability of these aspects within the system, and the human aspect (intelligence, insight and experience) of conducting the assessment itself. 

Some professionals were also wary of automation due to the numeric approaches that would be required to support it. These approaches, in their opinion, were subjective and highly dependent on the numbers input---numbers, which they note, are often precise but not necessarily accurate. This is a well-known shortcoming of quantitative approaches towards defining and assessing security risks~\cite{taylor2015potential}. 

Two other related challenges were mentioned by professions in using automated approaches for the IoT. The first was in determining appropriate levels of detail in which to conduct such an assessment and the second was deciding how best to combine detailed mathematical analyses on lower-level system components (data, devices, software and subsystems) to define an aggregated system-level risk. As aptly summarised by individuals:

\begin{quote}
	\textit{There's an issue with how much detail one goes into with a risk assessment. Actually risk assessment is about the aggregation of all of the individual risks and looking at the big picture.}
\end{quote}
	
\begin{quote}
	\textit{The real challenge is how to combine risks which are at different levels that relate to the same system, to determine the system-level risk. This is an area of active investigation. Mathematical analysis could be done, but this is difficult because accurate numbers are largely `unknown'.}
\end{quote}
 
According to professionals, these issues were key reasons why qualitative risk assessment methods were still preferred in industry. In those cases, discussions between risk officers and company personnel would determine which series of high-level categories (comparable to high, medium, low) to assign to threat likelihoods, impacts and resulting risks. This raises intriguing questions for the IoT given that a purely manual risk assessment is infeasible in such a large set of constantly changing system components.  

While participants were not keen on fully automated assessment, some admitted that there was scope for computer-assisted assessment approaches in connected systems. As noted:

\begin{quote}
	\textit{There is probably scope for computer-assisted risk assessment, for example, where a computer automatically maps the topology or devices on the network, their software, patch levels, etc. Also, the computer could help in qualitative assessments of risk, probability, impact, etc.}
\end{quote}

The features mentioned above are undoubtedly quite useful at supporting manual risk assessment, but whether they could adequately support highly dynamic IoT environments is still an outstanding question. These environments encourage flexibility and fluidity in how devices, systems and data are used in pursuit of fulfilment of the specified application scenario. 

\subsection{What role may collaborative risk assessment play?}

The last significant point that arose from the sessions pertained to collaborative risk assessment. In particular, some individuals felt that the current status quo of disjointed, internally-focused risk assessment would fail in IoT systems because of their coupling with external parties. Going forward, companies would therefore need to broaden the scope of assessments to incorporate business partners and other entities in their value chain. One participant commented:

\begin{quote}
	\textit{For future IoT risk assessments, this should be done across the supply chain, jointly if possible. The economy can get much more dynamic if there is a risk assessments process that allows action across the supply chain.}
\end{quote}

This proposal of a more largely scoped and collaborative risk assessment has value for numerous reasons. Firstly, it supports a comprehensive understanding of an IoT system, secondly, it allows a cross-enterprise appreciation of shared risk, and thirdly, it could allow for resources to be pooled and weaker (less security-inclined) organisations to be supported. 

A challenge to be faced however, would be the social element of organisations openly sharing their IoT risk data with partners outside of rigid legal contracts. The field of threat intelligence sharing has enumerated a host of the issues present (see~\cite{skopik2016problem}) as enterprises seek to work together for their joint security. Furthermore, collaborative risk assessment has been explored in the past (due to other technologies such as web services), but has had little success outside of tightly coupled (extended) enterprises~\cite{nurse2010business}. This therefore does not relate well to the IoT where coupling is dynamic and could be persistent or ephemeral. 

Instead of collaborative risk assessment, one professional expressed that organisations will need to view inter-organisational interactions and the risk they pose differently. His view was that:

\begin{quote}
	\textit{The more that organisations allow dynamic interactions between services, a point will be reached quite soon where they will have to assume that anything that goes out of the organisation is compromised and it doesn't matter how good the companies are, there will be a weak point on the chain. With these factors in mind, risk assessment will now need to be based on this knowledge. }
\end{quote}

This perspective is not unique and can be seen outside of the IoT---in a PwC report for instance, a similar concept was referred to as `operating in an assumed state of compromise'~\cite{pwc2013compr}. As mentioned by the individual, the main question becomes how should risk assessment be changed to incorporate this view. It is also worth noting that while such a view may suit security professionals, always assuming compromise (e.g., of data received from partnering entities) can have a significant impact on business efficiency. This is certain to raise issues when we consider that a prime reason for the adoption of the IoT is to increase efficient of services and applications.

\subsection{Implications of our findings}

Reflecting on our findings above, it is clear that there are several challenges and open questions regarding conducting risk assessments in the IoT environment. The evolving nature of the IoT is advantageous given how it adapts to our needs, however, it also renders many features of existing risk assessment approaches outdated. Instead of periodic assessments, more dynamic techniques will need to be explored to identify, evaluate and prioritise risks in such progressive systems. These techniques will also need to maintain a balance between dynamism, automation and human aspects, given the strong support for a social component to be present within risk assessment. Moreover, while the extent to which automated and mathematical methods are used is still an open question, these have their own challenges. In particular, there is the on-going difficulty in determining accurate values (about probability of attack or cost of attack impacts) when adopting such techniques to assess risks or risk components~\cite{taylor2015potential}. 

Another open question for the IoT pertains to the new series of risks that are emphasised as a result of it, and how risks can be assessed in such complex, automated, partially unknown systems. This requires joint efforts by academia and industry, both in researching potential approaches and trialling them on real systems. 
A good example of this is the logical glue that binds some IoT systems, and investigating how this may be incorporated into risk assessment, particularly if it is to be dynamic. If we consider the challenge of unknown systems, one way this may be addressed is by close collaboration and information sharing on risks between partnering businesses; i.e., collaborative assessment and expanded scopes of control and trust. This, however, is likely to be difficult. Furthermore, it may only work in cases where IoT systems across organisations are tightly and persistently coupled (much like those in extended enterprises~\cite{nurse2010business})---this is contrary to the fluid nature of the IoT.

There are social and legal challenges worthy of note as well. As IoT technologies become more ingrained into our world (from aviation to healthcare) and into business environments (offices to shop floors), a good security culture that appreciates the risks is essential. The criticality of such a culture is not new, but is further stressed in the IoT where security needs to be viewed as a continuous activity. This requires enhanced organisational awareness and training which appreciates how we as humans engage with security~\cite{bada2015cyber,furnell2017security}, and a progression from a periodic assessment mentality. 

While there are no specific regulations covering IoT security at this time (and questions about their broad feasibility in general~\cite{weber2016cybersecurity}), there are many legal issues that may arise. For instance, the European Union's (EU) General Data Protection Regulation (GDPR)~\cite{eugdpr2016} has strict requirements regarding the collection and processing of personal data. Organisations will therefore need to know exactly what data is collected by the IoT systems they have deployed, and ensure that risk assessments are suited to identify all associated risks. Another pertinent regulation is the NIS Directive which is an EU-wide legislation on cybersecurity~\cite{eunis2016}. This also places a number of other requirements on companies that will impact IoT systems, including how they are assessed, secured and managed. Future proposals for risk assessment in IoT systems will need to take account of all of these aspects and local legislation in the country of operations.

\section{Conclusion and future work}
\label{Conclusion and future work}

As the Internet of Things is adopted by more of society, the importance of understanding the risks that accompany it is paramount. In this paper, we sought to examine a number of key issues that pertain to identifying and assessing cyber-risk in IoT systems. Our aim was to validate and complement our earlier theory-driven research by gathering data and feedback from relevant industry professions through a series of workshops and interviews. These would provide useful insight and be based on real practitioner issues.

From an analysis of the data gathered, we confirmed several of the issues hypothesised, particularly the challenge of assessing risks in evolving IoT environments where knowledge of systems is constrained. \balance Other key topics that emerged included the perceived infeasibility of fully automated risk assessment in the IoT, and a view towards inter-organisational assessment of risk given IoT's wide connectivity. With an appreciation of these real-world challenges and industry insights, the next step in our research is to develop an enhanced risk assessment approach for the IoT. This would seek to address the issues defined while balancing the dynamic nature of the IoT with the rigour and structure of good risk assessment. 



\newcommand{\BIBdecl}{\setlength{\itemsep}{0.05 em}}
\bibliographystyle{IEEEtran}
\bibliography{bib}

\end{document}